\pdfoutput=1   
\documentclass[usenatbib,usegraphicx,fleqn]{mn2e}
\usepackage{amsmath,amssymb,bm,color,multirow,multicol}
\usepackage{charter}   

\newcommand{\e}[1]{\times 10^{#1}}
\newcommand{\fig}[1]{Fig. \ref{#1}}

\usepackage{ulem}

\title[Simulations of CTTS BP Tau]{Global 3D Simulations of Disc Accretion onto the classical T Tauri Star
BP Tauri}

\author[M. Long et al.]
{M. Long,$^1$\thanks{E-mail:long@astro.cornell.edu}, M. M. Romanova,$^2$\thanks{romanova@astro.cornell.edu},
 A. K. Kulkarni $^3$ and J.-F. Donati $^4$\\
$^1$ Center for Theoretical Astrophysics, Department of Physics, University of Illinois at Urbana-Champaign, Urbana, IL 61801-3080\\
$^2$ Department of Astronomy, Cornell University, Ithaca, NY 14853-6801, USA\\
$^3$ Harvard-Smithsonian Center for Astrophysics, 60 Garden Street, Cambridge, MA 02138, USA\\
$^4$ LATT-UMR 5572, CNRS\& Univ. P. Sabatier, 14 Av. E. Belin, F-31400 Toulouse, France}

\begin{document}



\maketitle


\begin{abstract}

\noindent Recent spectropolarimetric observations of the classical
T Tauri star BP Tau and analysis of its surface magnetic field
have shown that the magnetic field can be approximated as a
superposition of slightly tilted dipole and octupole moments with
respective strengths of the polar magnetic fields of 1.2 kG and
1.6 kG (\citealt{dona08}, hereafter D08). We adopt  the measured
properties of BP Tau and model the disc accretion onto the star by
performing global three-dimensional magnetohydrodynamic
simulations. We observed in simulations that the disc is disrupted
by the dipole component and matter flows towards the star in two
funnel streams which form two accretion spots below the dipole
magnetic poles. The octupolar component becomes dynamically
important very close to the star and it redirects the matter flow
to higher latitudes and changes the distribution and shape of the
accretion spots. The spots are meridionally elongated and are
located at higher latitudes, compared with the pure dipole case,
where crescent-shaped, latitudinally elongated spots form at lower
latitudes. The position and shape of the spots are in good
agreement with observations.

The disk-magnetosphere interaction leads to the inflation of the
field lines and to the formation of magnetic towers above and
below the disk. The magnetic field of BP Tau is close to the
potential inside the magnetospheric surface, where magnetic stress
dominates over the matter stress. However, it strongly deviates
from the potential at larger distances from the star.

A series of simulation runs were performed at different accretion
rates. In one of them, the disk is truncated  at $r\approx
(6-7)R_\star$ which is close to the corotation radius,
$R_{cor}\approx 7.5 R_\star$. However,  the accretion rate,
$1.4\times 10^{-9}M_\odot\mathrm{yr}^{-1}$, is lower than that
obtained from most of the observations. In a sample model with a
higher accretion rate $8.5\times 10^{-9}M_\odot\mathrm{yr}^{-1}$,
the disk is truncated at $r\approx 3.6R_\star$, but such a state
can not be a typical state for the slowly rotating BP Tau if it is
in the rotational equilibrium. However, torque acting on the star
is also small: it is about an order of magnitude lower than that
which is required for the rotational equilibrium. We suggest that
a star could lose most of its angular momentum at earlier stages
of its evolution.

\end{abstract}


\section{Introduction}

Accretion-powered Classical T Tauri stars (CTTSs) are young low-mass stars which
often show signs of a strong magnetic field (e.g., \citealt{basr92,john99})
 which is expected to have a complex structure (e.g., \citealt{john07}).
The Zeeman-Doppler imaging technique has proven very successful in obtaining
surface magnetic maps for many stars, and the external magnetic fields of the
stars have been reconstructed from these maps under the potential approximation
\citep{dona97, dona99, jard02, jard06, greg10}. The magnetic field plays a crucial role in disc
accretion by disrupting the inner regions
of the disc and channeling the matter onto the star, and hence it is important to know the
magnetic field configurations in magnetized stars.

D08 recently observed the CTTS BP Tau with the ESPaDOnS and NARVAL
spectropolarimeters and reconstructed the surface magnetic field
from the observations. They have shown that the magnetic field of
BP Tau can be approximated by a combination of  dipole and
octupole components of 1.2 kG  and 1.6 kG, which are slightly (but
differently) tilted about the rotational axis.

D07  analyzed the distribution of the accretion spots on the
stellar surface and found spots at high latitudes, which cover
about 8 per cent of the stellar surface (D08).

Further, D08 extrapolated the surface magnetic field to larger distances using the potential approximation,
(i.e., assuming that  there are no currents outside the star and hence the
external matter does not influence the initial configuration of the field)
and estimated the distance at which
the disk should be disrupted by the magnetosphere so that the matter flowing
towards the star in funnel streams
produces the high-latitude spots. They concluded that this distance should be
quite large, $r\gtrsim 4 R_\star$.  However, this problem requires more complete analysis
based on the MHD approach, where external currents can be taken into account, and the matter
flow around the magnetosphere can be investigated self-consistently, taking into account
interaction of the external plasma with the magnetic field.

In this paper, we investigate this problem using global three-dimensional MHD simulations.
 We solve the 3D MHD equations numerically in our simulation model to
 investigate the structure of the external magnetic field, accretion flows and
 location of accretion spots.

In our previous work, we have performed global 3D simulations of
accretion onto stars with misaligned dipole fields \citep{roma03,
roma04a, kulk05}, and  aligned or misaligned dipole plus
quadrupole fields \citep{long07,long08}. Recently, we were able to
extend our method and to build a numerical model for stars with an
octupolar component. The general properties of the model have been
described in detail in \citet{long10} and the model was  applied
to another CTTS, V2129 Oph, with a strong octupole field in
\citet{roma10} which has been compared with observations of V2129
(\citealt{dona07}, \citealt{dona10}).

In this paper, we apply our 3D MHD model of stars with complex
magnetic fields \citep{long10} to the CTTS BP Tau. We investigate
disc accretion onto the star by adopting the measured surface
magnetic fields (D08) and other suggested properties of this star.
The surface magnetic field is modeled as a superposition of a 1.2
kG dipole and 1.6 kG octupole field, tilted by $20^\circ$ and
$10^\circ$ with respect to the rotational axis and located at
opposite phases (the phase difference is $180^\circ$). We also
take into account other parameters of  BP Tau: its mass
$M_\star=0.7 M_\odot$ \citep{sies00}, and radius $R_\star=1.95
R_\odot$ \citep{gull98}. Its age is about 1.5 Myr (D08), and its
rotation period is $7.6$ days \citep{vrba86}, which corresponds to
a corotation radius of $R_{cor}\approx7.5 R_\star$.
The mass accretion rate derived from different observations varies
between $\dot{M}\simeq2.9\times10^{-8}$M$_\odot$yr$^{-1}$ (e.g.
\citealt{gull98}) and $9\times10^{-10}$M$_\odot$yr$^{-1}$
\citep{schm05}. We performed a series of simulation runs at
different mass accretion rates in order to investigate the cases
where the disk stops at different distances from the star. We also
calculated the torque on the star and compared it with star's age.

To understand the role of the octupole field in channeling the
accreting matter, we compared our dipole plus octupole model of BP
Tau with a similar model but with only the dipole component.

Thus, we focus on: (1) 3D MHD modeling of accretion flows around
CTTS BP Tau modeled with dipole plus octupole moments; (2)
comparisons of accretion properties observed in simulations with
observations of BP Tau, such as the shape and distribution of hot
spots, mass accretion rates and more; (3) deviation of the
simulated magnetic field from the potential field.

Section \S2 briefly describes the numerical model used in this work.
The simulation results are shown in \S 3.
We end in \S4 with our conclusions and some discussion.

\section{Model}

The global 3D MHD model originally developed by \citet{kold02} and
used for modeling stars with dipole fields \citep{roma03,roma04a}
was modified to include higher order components
\citep{long07,long08,long10,roma10} to simulate disc accretion
onto stars with complex fields. The MHD equations are solved in a
reference frame co-rotating with the star. A viscous term is
incorporated into the MHD equations (only in the disc) to control
the rate of matter flow through the disk. We use the
$\alpha-$prescription for viscosity with $\alpha=0.01$.

\begin{enumerate}

\item

\textit{Initial conditions.} The simulation domain consists of a cold, dense disc and a hot,
low-density corona, which are initially in rotational hydrodynamical equilibrium.
The initial angular velocity in the disc is close to  Keplerian.
The angular velocity in the corona at any given cylindrical radius is
set to be equal to that of the disk at that radius.

\item

\textit{Boundary conditions.} At the inner boundary (the surface of the star),
most of the variables $A$ are set to have free boundary conditions, ${\partial A}/{\partial r}=0$.
The initial magnetic field on the surface of the star is taken to be a
superposition of misaligned dipole and octupole fields. As the simulation proceeds,
we assume that the normal component of the field remains unchanged, i.e.,
the magnetic field is frozen into the surface of the star. At the outer boundary,
free conditions are taken for all variables. In addition, matter is not
permitted to flow into the region from the outer boundary.

\begin{table}
\caption{The reference values for CTTS BP Tau. The dimensional
values can be obtained by multiplying the dimensionless values
from simulations by these reference values. $B_0$ and the
subsequent values below depend on $\widetilde\mu_1$.} \centering
\begin{tabular}{llll}
\hline
Reference Units          & $\widetilde{\mu}_1=1$ & $\widetilde{\mu}_1=2$ &  $\widetilde{\mu}_1=3$                \\
\hline
{$M_\star(M_\odot)$}              & $0.7$          & --  &  --         \\
{$R_\star(R_\odot)$}              & $1.95$         & --  &  --       \\
{$B_{1\star}$ (G)}                & $1200$         & --  &   --          \\
{$R_0$ (cm)}                      & $3.9\e{11}$    & --  &   --          \\
{$v_0$ (cm s$^{-1}$)}             & $1.5\e7$       & --  &   --       \\
{$P_0$ (days)}                    & $1.83$         & --  &   --          \\
{$B_0$ (G)}                       & $25.7$          & $12.9$  & $8.6$                 \\
{$\rho_0$ (g cm$^{-3}$)}          & $2.8\e{-12}$   & $6.9\e{-13}$ & $2.8\e{-13}$   \\
{$\dot M_0$ ($M_\odot$yr$^{-1}$)} & $1.0\e{-7}$    & $2.6\e{-8}$  & $1.1\e{-8}$    \\
{$F_0$ (erg  cm$^{-2}$s$^{-1}$)}  & $1.0\e{10}$    & $2.6\e{9}$   & $1.1\e{9}$     \\
{$N_0$ (g cm$^2$s$^{-2}$)}  &   $3.9\e{37}$  &  $9.8\e{36}$  & $4.6\e{36}$       \\
\hline
\end{tabular}
\label{tab:refval}
\end{table}

\item

\textit{Simulation region and grid.} We use  the ``cubed sphere"
grid introduced by \cite{kold02} (see also Fig. 1 in
\citealt{long10}). The resolution of the grid is
$6\times120\times51^2$ to simulate the accretion onto BP Tau in a
simulation domain of $41R_\star$. We use a very high resolution
near the star in order to resolve the complex structure of the
octupolar component of the field.

\item

\textit{Magnetic field configuration.} In our code we can model the magnetic field of the star
by a superposition of three
multipole moments $\bm\mu_i$ ($i=1,2,3$ for dipole, quadrupole and octupole respectively)
which are tilted relative to the $z-$ axis (which is aligned with the rotational axis  $\bm\Omega$)
 at different angles $\Theta_i$, and have different angles $\phi_i$ between the $xz$ plane and
$\bm\Omega-\bm\mu_i$ planes. For simplicity, $\phi_1$ is set to be
0, which means that the dipole moment $\bm\mu_1$ is in $xz$ plane.
The general magnetic field configuration is discussed in greater
detail  in \citet{long10}. The details of the magnetic field
configuration in our
 BP Tau model are discussed in \S 3.1.

\item

\textit{Reference units.}
The simulations are performed in dimensionless variables $\widetilde{A}=A/A_0$ where $A_0$ are reference values.
We choose the stellar mass $M_\star$, radius $R_\star$
and the surface dipole field strength $B_{1\star}$ to build a set of reference values.
The reference values are: length scale: $R_0=R_\star/0.35$; velocity: $v_0=(GM_\star/R_0)^{1/2}$;
 time-scale: $P_0=2\pi R_0/v_0$.
The reference magnetic moments for dipole and octupole components
are $\mu_{1,0}=B_0R_0^3$  and $\mu_{3,0}=B_0R_0^5$ respectively,
where $B_0$ is the reference magnetic field. Hence, the
dimensionless magnetic moments are:
$\widetilde{\mu}_1=\mu_1/\mu_{1,0}$,
$\widetilde{\mu}_3=\mu_3/\mu_{3,0}$, where the dipole and octupole
moments $\mu_1$ and $\mu_3$ of the star are fixed. We take one of
the above relationships; for example, the one for the dipole; to
obtain

\begin{equation}
\label{eq-B0} B_0=\frac{\mu_{1,0}}{R_0^3} = \frac{0.5
B_{1\star}}{\widetilde{\mu_1}}\bigg(\frac{R_\star}{R_0}\bigg)^3=
25.7\bigg(\frac{B_{1\star}}{1.2 {\rm
kG}}\bigg)\bigg(\frac{1}{\widetilde{\mu_1}}\bigg) {\rm G} .
\end{equation}
 Hence, at fixed $B_{1\star}$,  the reference magnetic field depends on
the dimensionless parameter $\widetilde\mu_1$. The reference
density $\rho_0$ and the mass accretion rate $\dot{M}_0$ also
depend on this parameter:
\begin{equation}
\label{eq-rho0} \rho_0=B_0^2/v_0^2 = 2.8\times10^{-12}
\bigg(\frac{B_{1\star}}{1.2 {\rm
kG}}\bigg)^2\bigg(\frac{1}{\widetilde{\mu_1}}\bigg)^2 \frac{\rm
g}{{\rm cm}^3} ,
\end{equation}
\begin{equation}
\label{eq-Mdot0} \dot{M}_0=\rho_0 v_0
R_0^2=1.0\times10^{-7}\bigg(\frac{B_{1\star}}{1.2 {\rm
kG}}\bigg)^2\bigg(\frac{1}{\widetilde{\mu_1}}\bigg)^2
\frac{M_\odot}{{\rm yr}}.
\end{equation}

The dimensional accretion rate then is $\dot M=\widetilde{\dot
M}\dot M_0$, where $\widetilde{\dot M}$ is the dimensionless
accretion rate.
 One can see that
the dimensional accretion rate depends on $\widetilde{\dot M}$
which we find from simulations, and the dimensionless parameter
$\widetilde\mu_1$, which we use to vary the accretion rate. We
change the dimensionless octupolar moment $\widetilde\mu_3$ in
same proportion so as to keep the ratio $\mu_3/\mu_1$ fixed. To
find the ratio between the dimensionless moments, we use
approximate formulae for aligned moments: $\mu_1=0.5 B_{1\star}
R_\star^3$ and $\mu_3=0.25 B_{3\star} R_\star^5$  (see
\citealt{long10}) and obtain for BP Tau the ratio
$\widetilde{\mu}_3/\widetilde{\mu}_1=B_{3\star} \tilde R_\star^2/2
B_{1\star}\approx 0.08$ (where $\tilde R_\star=0.35$).

Other reference values are: angular momentum flux (a torque)
$\dot{N}_0=\rho_0v_0^2R_0^3$; energy flux
$\dot{E}_0=\rho_0v_0^3R_0^2$; temperature
$T_0=\mathcal{R}p_0/\rho_0$, where $\mathcal{R}$ is the gas
constant; and the effective blackbody temperature
$T_{\mathrm{eff,0}} = (\rho_0 v_0^3/\sigma)^{1/4}$, where $\sigma$
is the Stefan-Boltzmann constant. Tab. \ref{tab:refval} shows the
reference values for CTTS V2129 Oph. In the subsequent sections,
we show dimensionless values $\widetilde{A}$ for most of the
variables and drop the tildes($\sim$). However, we keep them in
 $\widetilde\mu_1$, $\widetilde\mu_3$, and  $\widetilde{\dot M}$ because these are important parameters of the model.

\item

\textit{The magnetospheric radius.} The truncation radius, $r_t$,
where the disc is truncated by the magnetosphere,
 could be estimated as (e.g., \citealt{elsn77}):
\begin{equation}
r_t=k(GM_\star)^{-1/7}\dot{M}^{-2/7}\mu_1^{4/7},
\end{equation}
where $\dot M$ is the accretion rate and $\mu_1$ is the dipole magnetic moment; $k$
is a coefficient of  order unity. For example, \cite{long05} obtained
$k\approx 0.5$ in numerical modeling of  disc-accreting stars.
For stars with known $M_\star$, $R_\star$, and dipole magnetic moment $\mu_1$,
such as BP Tau, the accretion
rate determines where the disc stops and how the matter flows onto the star.

\end{enumerate}

\section{Modeling of accretion onto BP Tau}


We performed  a number of simulation runs at different accretion
rates and observed that the disk is truncated at different radii.
We choose as our main case the one in which the disk stops
sufficiently far away, but at the same time the accretion rate is
not very low.
 Below, we describe  this case in detail.
 We also show the results
at higher accretion rates for comparisons (see \S 3.7).

\subsection{The modeled magnetic field}

D08 decomposed  the observed surface magnetic field of BP Tau into
spherical harmonics and found  that the field is mainly poloidal with
only $10\%$ of the total magnetic energy in the toroidal field.
The poloidal component can be approximated by
dipole ($l=1$) and octupole ($l=3$) moments with $50\%$ and $30\%$
of the magnetic energy respectively. Other multipoles (up to $l<10$) have
only $10\%$ of the total magnetic energy. D08 concluded that the magnetic field of BP Tau
is dominated by a 1.2 kG dipole and 1.6 kG octupole tilted by $20^\circ$
and $10^\circ$.
The meridional angle between the $\bm\Omega-\bm\mu_1$ and
$\bm\Omega-\bm\mu_3$ planes is approximately $180^\circ$. In our model, we only consider the poloidal component.

We convert the above parameters into  dimensionless values using
our reference units and solve 3D MHD equations (see, e.g.,
\citealt{kold02}), in  dimensionless form. One of the important
parameters of the model is $\widetilde\mu_1$ which is used to vary
the truncation radius in the dimensionless model (and the
accretion rate in the dimensional model, see eq. 1). We find the
second dimensionless parameter $\widetilde\mu_3$ from the
relationship: $\widetilde{\mu}_3/\widetilde{\mu}_1=B_{3\star}
\tilde R_\star^2/2 B_{1\star}\approx 0.08$ (where $\tilde
R_\star=0.35$). This ratio is fixed for fixed values of the dipole
and octupole components, $B_{1\star}=1.2$kG and
$B_{3\star}=1.6$kG. From a number  of simulation runs at different
$\widetilde\mu_1$, we choose $\widetilde\mu_1=3$,
$\widetilde\mu_3=0.24$ to obtain a large enough magnetosphere as
suggested by D08 and investigate this case in detail. Other
parameters of the magnetic field configuration of BP Tau are
$\Theta_1=20^\circ$, $\Theta_3=10^\circ$ and $\phi_3=180^\circ$.

\begin{figure}
\begin{center}
\includegraphics[width=8.0cm]{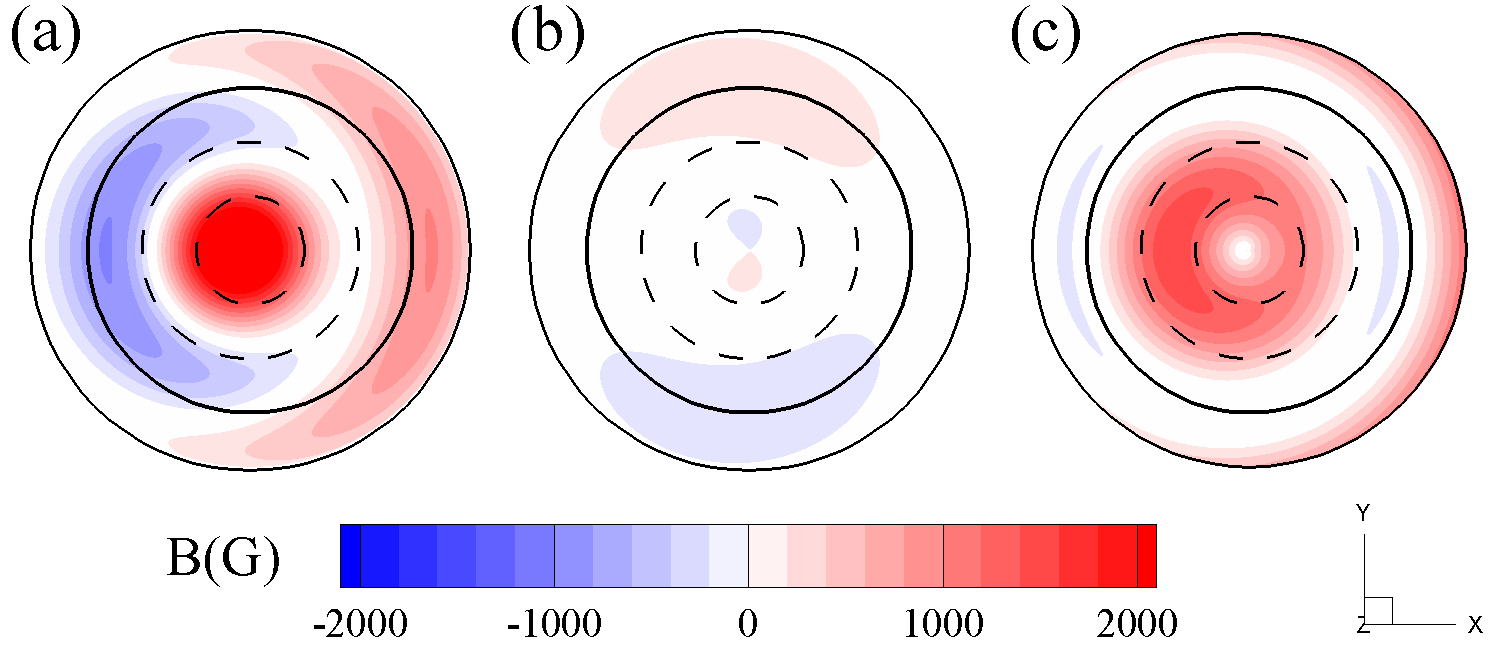}
\caption{\label{bcompb} : Polar projections of the magnetic field
components at the stellar surface in
the dipole plus octupole model of BP Tau in spherical
coordinates: radial magnetic field, $B_r$; azimuthal magnetic field, $B_\phi$;
and meridional magnetic
field, $B_\theta$.  The outer boundary, the bold circle and the two inner
dashed circles represent the latitude of $-30^\circ$, the equator, and the latitudes
of $30^\circ$ and $60^\circ$ respectively.  The red and blue regions represent
positive and negative polarities of the magnetic field.}
\end{center}
\end{figure}

\begin{figure}
\begin{center}
\includegraphics[width=8.0cm]{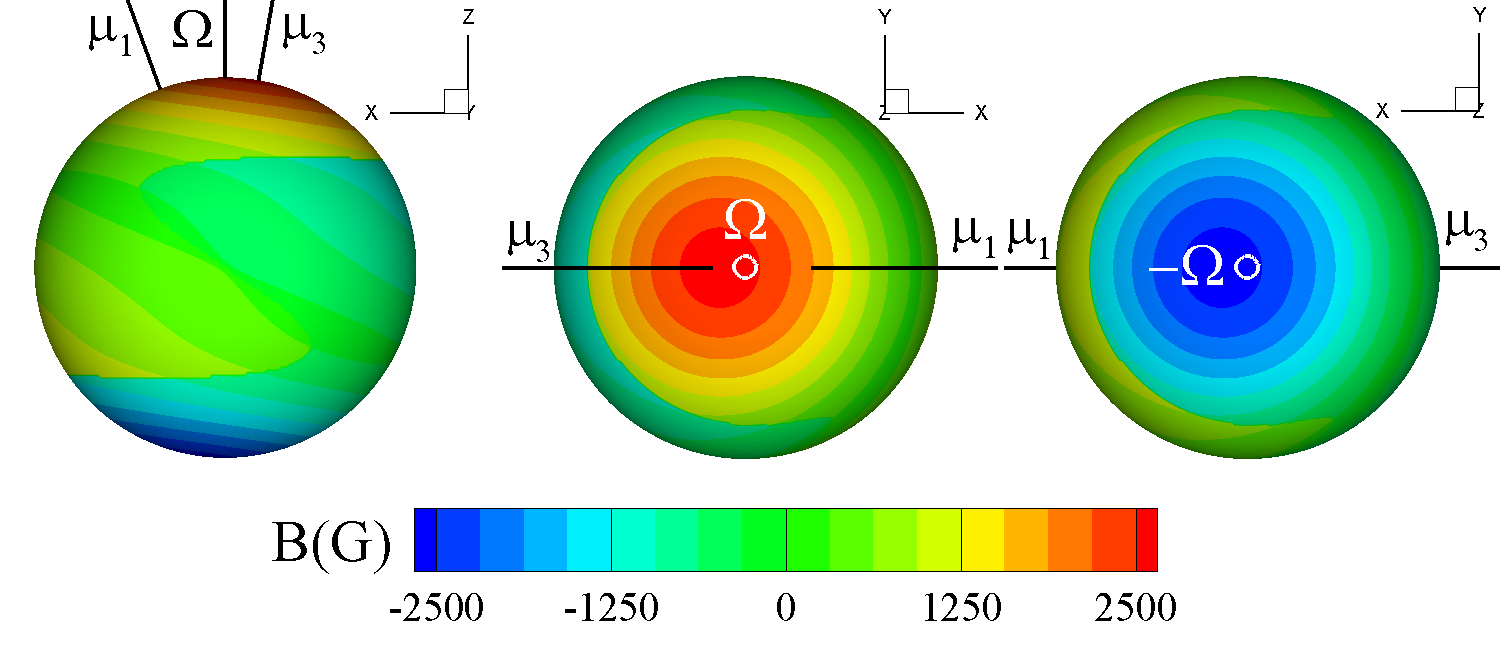}
\caption{\label{bsurfb} The surface magnetic field in the dipole
plus octupole model of BP Tau  ($\widetilde\mu_1=3$,
$\widetilde\mu_3=0.24$, $\Theta_1=20^\circ$, $\Theta_2=10^\circ$,
$\phi_3=180^\circ$) as seen from the equatorial plane (left
panel), the north pole (middle panel) and the south pole (right
panel). The colors represent different polarities
 and strengths of the
magnetic field.}
\end{center}
\end{figure}

\fig{bcompb} shows the components of the simulated magnetic field ($B_r$, $B_\phi$
and $B_\theta$) in the polar projection down to the latitude $-30^\circ$. One can see that
the distribution of the radial component
(left panel) is very similar to that obtained by D08 in  two observational epochs of Dec06 and Feb06
(see Fig. 14, left panels of D08).
In both cases, there is a strong positive pole at colatitudes of $0-30^\circ$,
a part of the negative octupolar belt at colatitudes of
$60^\circ-90^\circ$ and a part of the positive octupolar belt at colatitudes of
$90^\circ-120^\circ$. The distribution of the meridional component, $B_\theta$,
(right panel of \fig{bcompb}) is qualitatively similar to that of D08, though
in the D08 plot the inner positive ring (red color in the plot)
is weaker, while the negative ring (blue) is stronger compared with our model.
The azimuthal component of the field is weak
and shows a butterfly pattern of the negative and positive polarities
in both the modeled and reconstructed fields.

\fig{bsurfb} shows three-dimensional views of the magnetic field
distribution at the surface of the star. It can be seen that there
are two antipodal polar regions of opposite polarity where the
field is strongest. They  approximately coincide with the
octupolar high-latitude poles and their centers are located close
to the octupolar axis $\bm\mu_3$. Next to these regions, there are
negative (blue) and positive (red) octupolar belts. Their shapes
are more complex than those of the belts in pure octupole cases,
where the belts are parallel to the magnetic equator (see
\citealt{long10}). This is because the dipole component strongly
distorts the ``background" octupolar field.

Although the octupole field is strongest at the surface of the
star, it decreases more rapidly than the dipole field with
distance from the star. To investigate the role of the dipole and
octupole components in  channeling the accretion flow, we find the
radius at which the dipole and octupole fields are equal.  For
this, we assume that both magnetic moments are aligned with the
rotational axis and take the magnetic field in the magnetic
equatorial planes: $B_1=\mu_1/r^3$ and $B_3=3\mu_3/2r^5$ (see Eqn.
1 in \citealt{long10}). Noting that the strengths of the field at
the magnetic poles are $B_{1\star}=2\mu_1/R_\star^3$ and
$B_{3\star}=4\mu_3/R_\star^5$, and equating $B_1$ to $B_3$, we
find this radius:

\begin{equation}
r_{eql}=\left(\frac{3}{4}\frac{B_3\star}{B_1\star}\right)^{1/2}R_\star.
\end{equation}
Substituting $B_{1\star}=1.2$kG and $B_{3\star}=1.6$kG, we obtain $r_{eql}\approx R_\star$.
Note that both dipole and octupole moments are tilted about the rotational axis, and hence
the above formula gives only an approximate value for this radius. We suggest
that this radius should be located slightly above the surface of the star to explain the dominance of
the octupolar field seen in \fig{bcompb} and \fig{bsurfb}.

\begin{figure*}
\centering
\includegraphics[width=14.0cm]{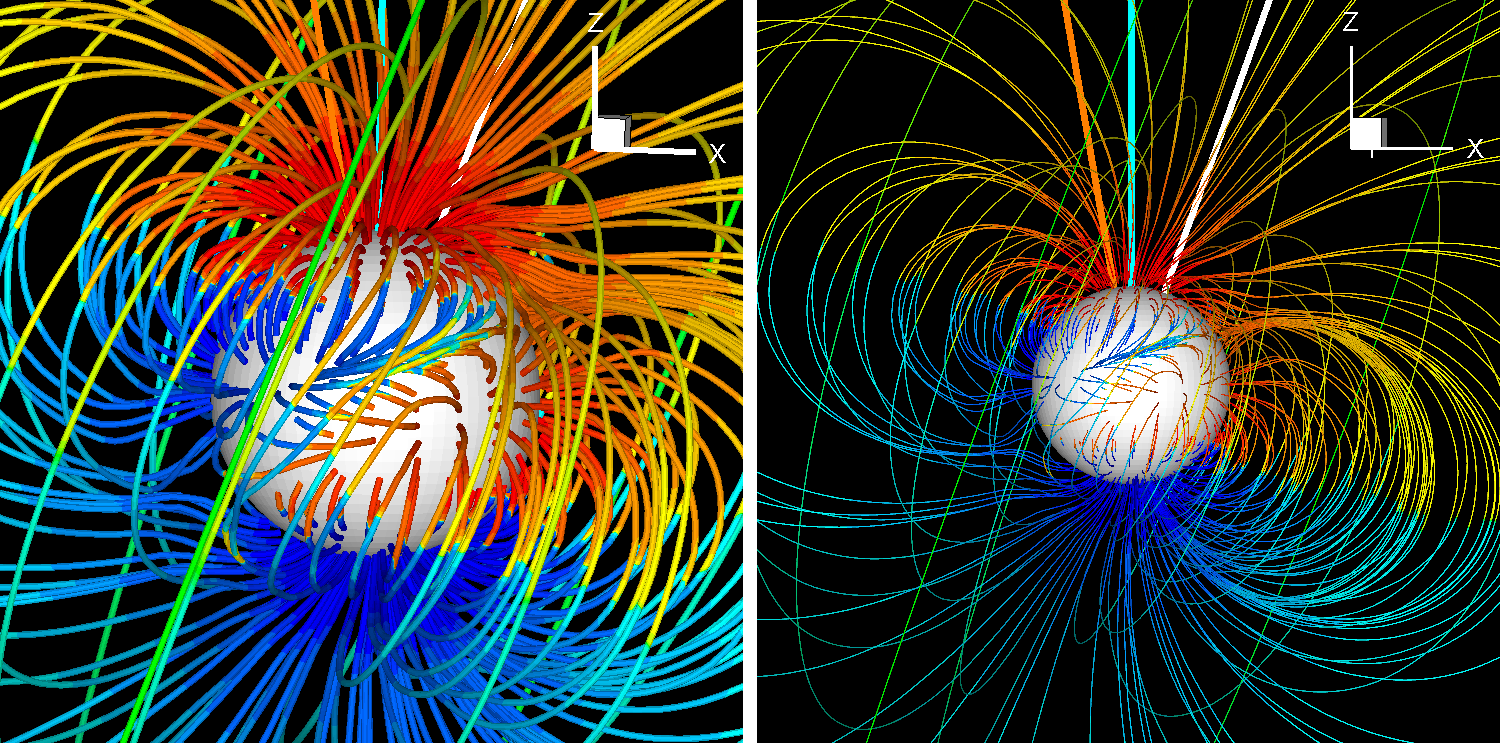}
\caption{\label{magb} The magnetic field of BP Tau modeled as a
superposition of dipole and octupole components at $t=0$. The
field shows an octupolar structure only very close to the star,
while the dipole field dominates in the rest of the simulation
region. The color of the field lines represents the polarity and
strength of the field. The thick cyan, white and orange lines
represent the rotational axis and the dipole and octupole moments
respectively.}
\end{figure*}

\fig{magb} shows the initial magnetic field distribution near the star in our model.
 One can see  the octupolar field component in the vicinity of the star. It also
modifies the dipole field  up to distances of
 $r\approx 0.5 R_\star$ (above the surface of the star). The dipole field dominates in the rest of the simulation region.
 We should note that the dipole and octupole fields are equal at $r_{eql}\approx R_\star$, but
 the octupole component disturbs the dipole field up to larger distances.

\subsection{Matter flow and the magnetic field structure}

\begin{figure}
\begin{center}
\includegraphics[width=8.5cm]{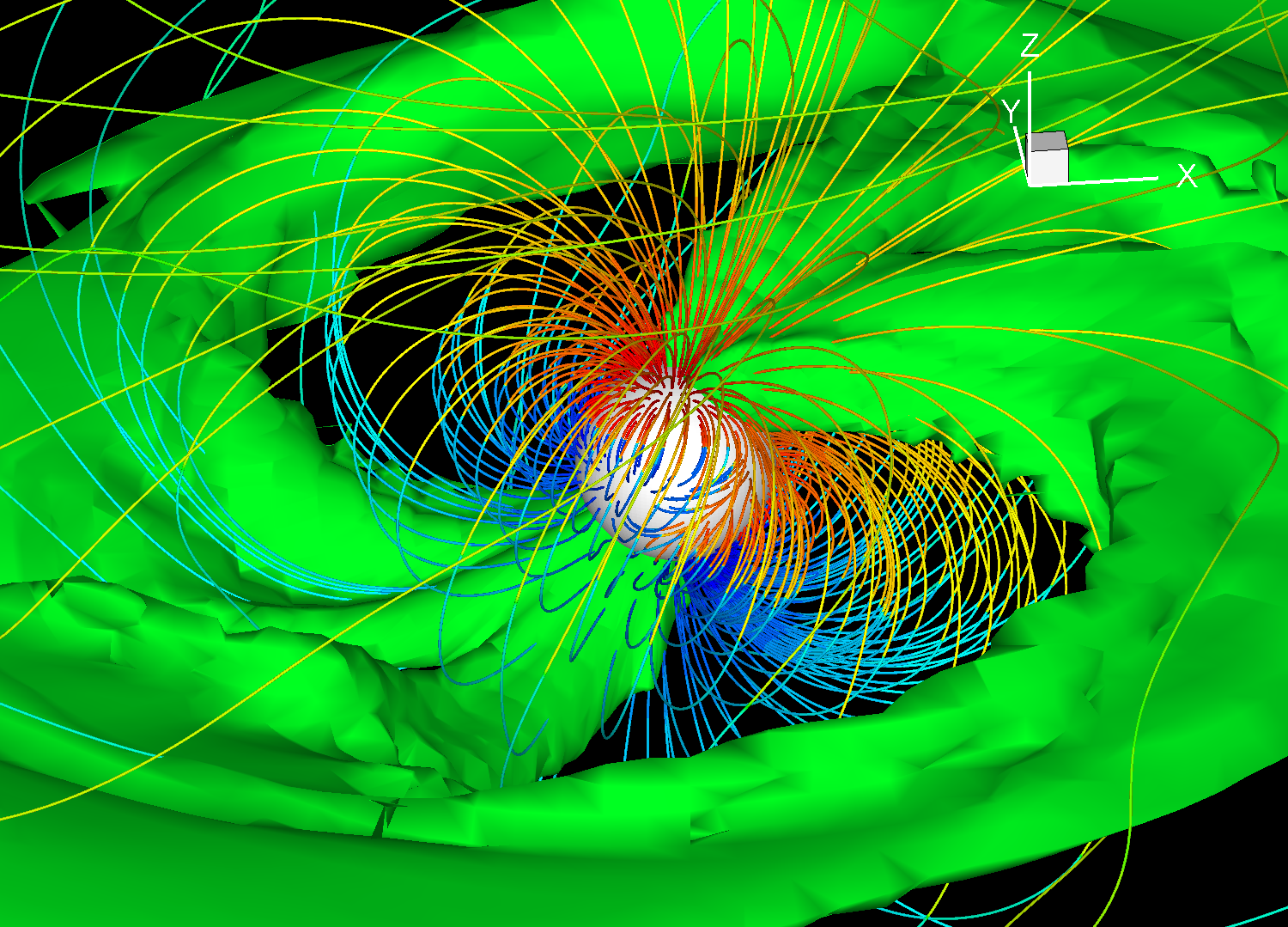}
\caption{\label{3db} 3D view of matter flow and the magnetic field
distribution in the main case at $t=10$. The left panel shows the
distribution of the magnetic field lines. The middle panel shows
one of the density levels, $\rho=0.84\times10^{-13}$. The right
panel shows the density distribution in the disc plane. The colors
along the field lines represent different polarities and strength
of the magnetic field. The thick cyan, white and orange lines
represent the rotational axis and the dipole and octupole moments
respectively.}
\end{center}
\end{figure}

Here, we show results of 3D MHD simulations of matter flow onto the dipole plus octupole model
of BP Tau with parameters corresponding to the main case
($\widetilde\mu_1=3$, $\widetilde\mu_3=0.24$) at time $t=10$ when the system is in a
quasi-stationary state. Fig. \ref{3db} (middle panel) shows that the disk is truncated by the dipole component of the field, and matter flows
towards the star in two ordered funnel streams. The right panel shows the density distribution in
the equatorial plane.
The left panel shows that the magnetosphere is disturbed by the disk-magnetosphere interaction.

\begin{figure}
\begin{center}
\includegraphics[width=8.5cm]{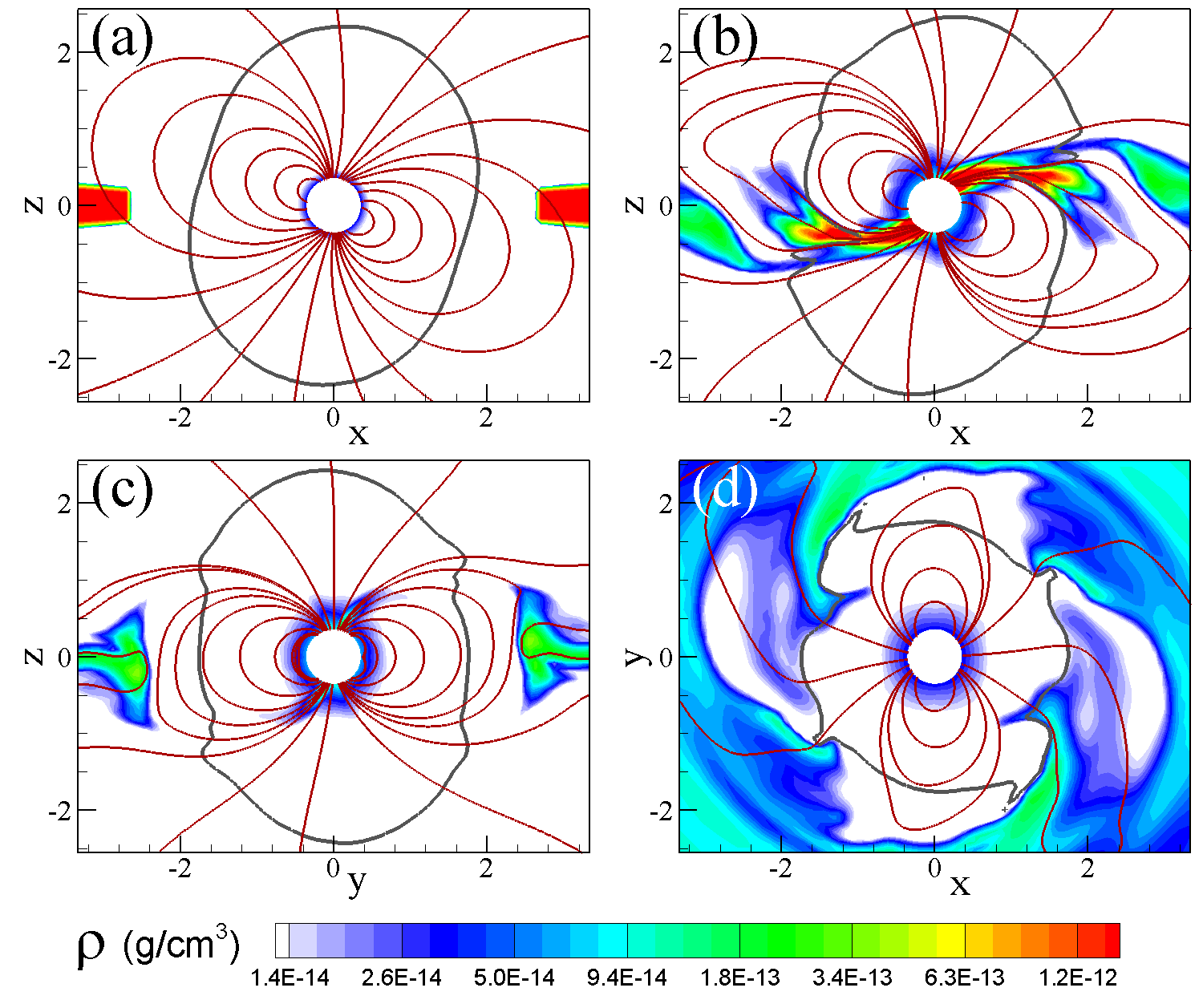}
\caption{\label{flowb} Density distribution and selected field lines in different slices.
Panels (a) and (b) show $xz$ slices
at $t=0$ and $t=10$. Panels (c) and (d) show $yz$ and $xy$ slices at $t=10$.
The gray line shows the distance at which the matter stress equals the magnetic stress.}
\end{center}
\end{figure}

\fig{flowb} shows slices of the density distribution and projected
field lines. Panel (b) shows the slice in the $\bf{\mu_1} -
\bf{\Omega}$ (or $xz$) plane. One can see more clearly that the
dipole component of the field is responsible for disk truncation
and matter is channeled in the $\bf{\mu_1} - \bf{\Omega}$ plane.
Panel (c) shows that  the disk is stopped by the magnetosphere in
the $yz-$plane. It is stopped at $r\approx (6-7)R_\star$. Panel
(d) shows that matter flow is complex in the equatorial plane, and
that matter comes closer to the star in the $yz-$plane.

The gray line in \fig{flowb}  shows the magnetospheric radius,
$r_m$, the distance where the matter and magnetic stresses are
equal: $\beta = (p+\rho v^2)/(B^2/8\pi) = 1$. The magnetic stress
dominates at $r<r_m$ (\citealt{ghosh78, ghosh79a, long10}). Fig.
\ref{flowb} shows that $r_m \approx 5R_\star$ which is smaller
than the actual truncation radius $r_t\approx (6-7) R_\star$. We
should note that there are different criteria for the truncation
radius and the above criterion usually gives the smallest
truncation radius (see \citealt{bess08} for details).

\begin{figure*}
\begin{center}
\includegraphics{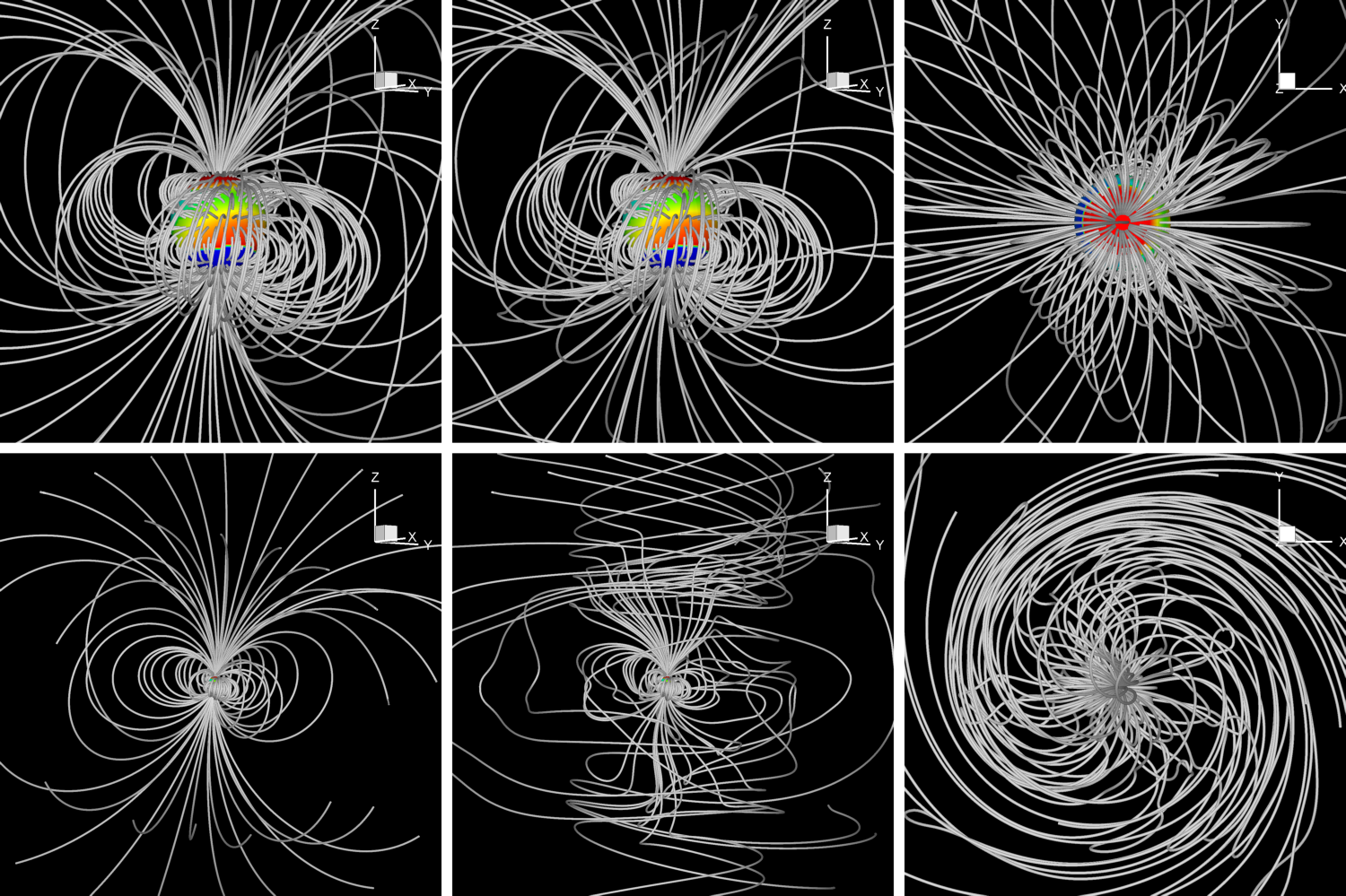}
\caption{\label{blinesb}
Comparison of the initial (potential) magnetic field distribution
at $t=0$ (left panels) with the field distribution at $t=10$ (middle and right panels).
The top panels show the magnetic field in the vicinity of the star, while the bottom panels
show the field distribution in the whole simulation region.
The left and middle columns show the side view, while the right column shows the axial view of the field
The color on the star's surface shows
different polarities and strengths of the magnetic field.}
\end{center}
\end{figure*}

The external magnetic field evolves and deviates from the initial configuration
due to disk-magnetosphere interaction. The left panels of
\fig{blinesb} show the initial field distribution at $t=0$
which corresponds to the potential field produced by the star's internal currents.
As the magnetic field evolves, the currents produced by the motion of plasma
outside the star become important and they change the potential magnetic field
significantly.
The middle and right panels show the evolved field from different directions at $t=10$.
It can be seen that on a large scale the magnetic field lines wrap around
the rotational axis of the star and form a magnetic tower (e.g. \citealt{lynd96, kato04, roma04b}).
This is a natural result of the magnetic coupling between the disk and star.
The foot-points of the field lines on the star rotate faster than the
foot-points threading the disk, which leads to  differential rotation along the
lines, and their stretching and  inflation into the corona. The rotational
energy is converted into the magnetic energy associated with these field
lines. The evolved field structure on the large scale significantly differs from
the potential field shown in the left panels.

In the vicinity of the star the situation is different. The middle top panel of
\fig{blinesb}
shows that the magnetic field distribution at $t=10$ is almost identical to the initial field
distribution.
We conclude that the potential approximation is valid only within the parts of the magnetosphere where
the magnetic stress dominates, that is, at $r\lesssim 5 R_\star$ in our case.
The field distribution strongly departs from potential at larger distances from the star.

\subsection{Accretion spots in BP Tau and in the Model}

\begin{figure}
\begin{center}
\includegraphics[width=8.0cm]{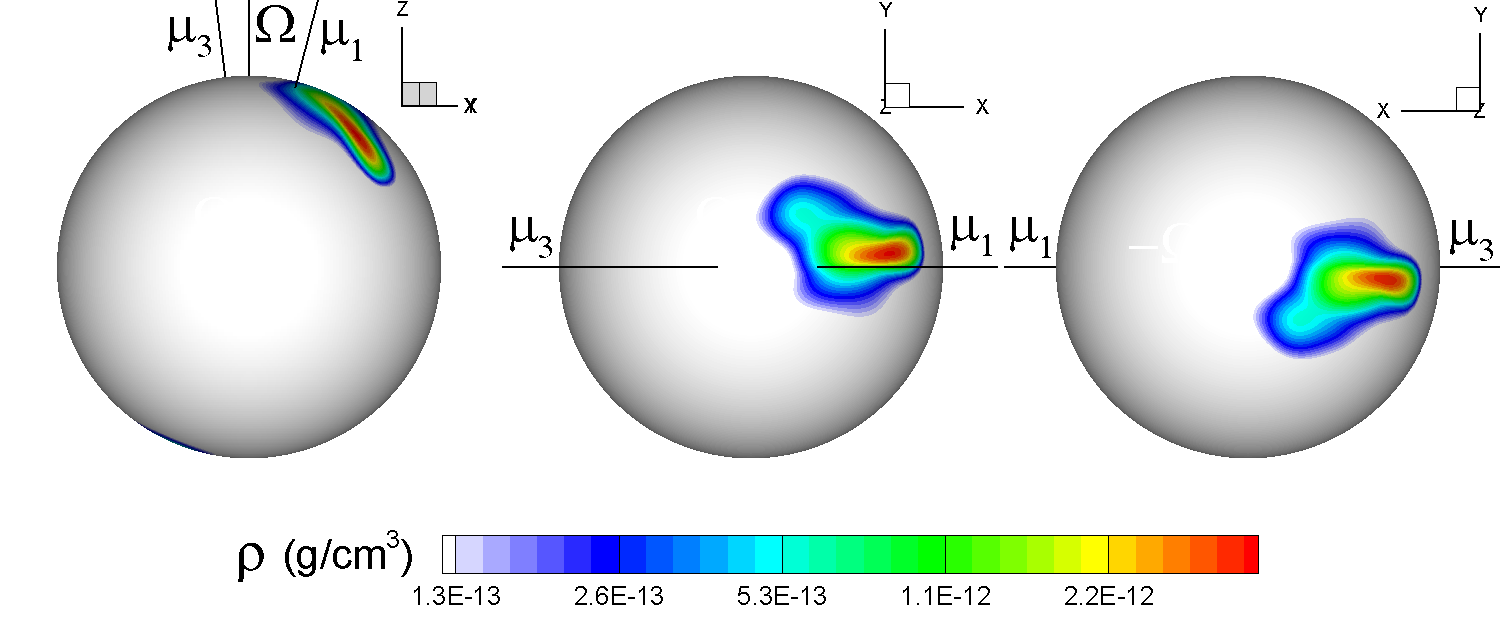}
\caption{\label{hsb}
The simulated accretion spots viewed from different directions
at $t=10$: from the equatorial plane
(left-hand panel), the north pole (middle panel), and the south pole
(right-hand panel). The color contours
show the density distribution of the matter. The solid lines represent the magnetic
moments of the dipole ($\bm{\mu_1}$) and octupole ($\bm{\mu_3}$).}
\end{center}
\end{figure}

\begin{figure}
\begin{center}
\includegraphics[width=6.0cm]{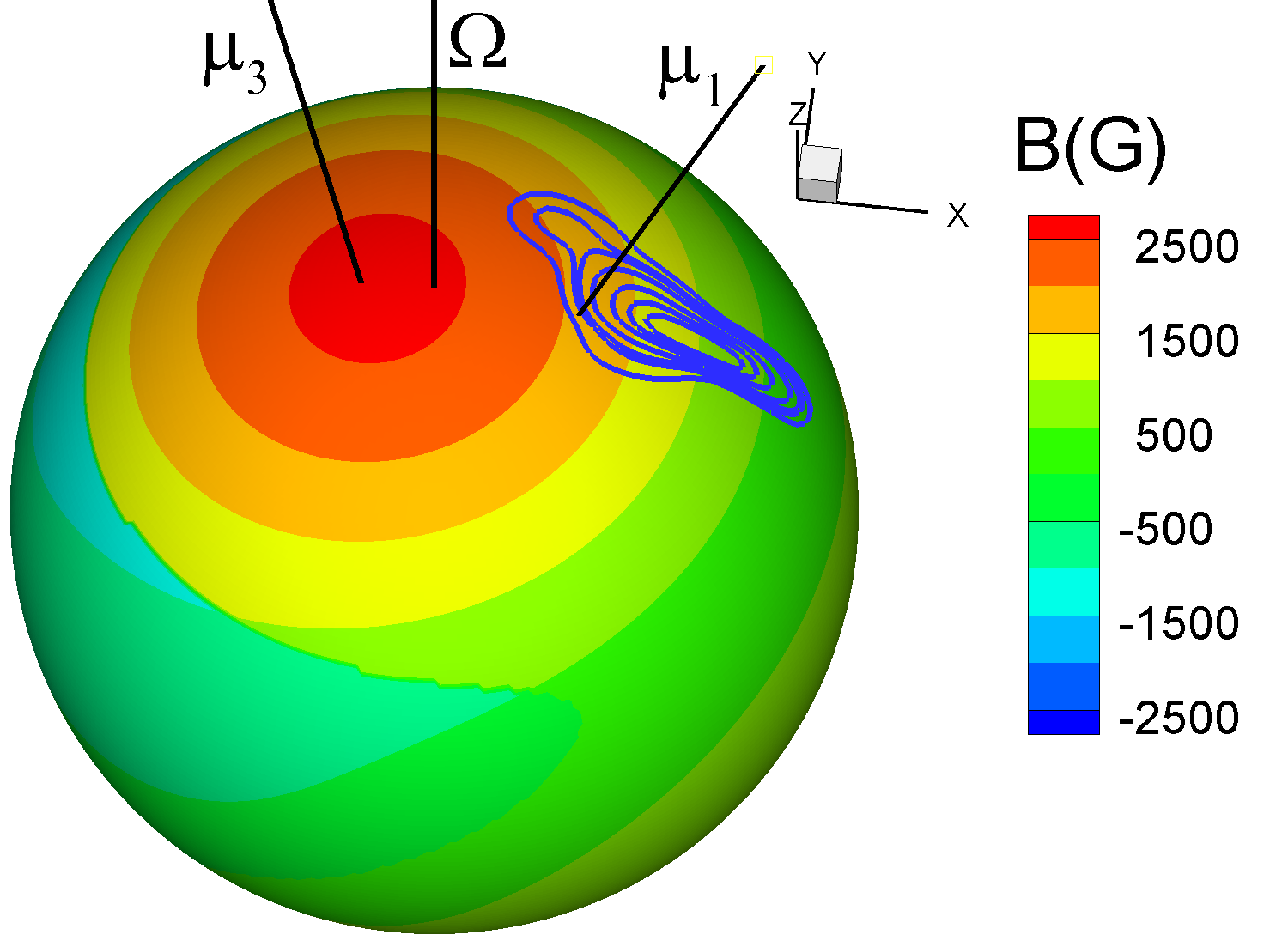}
\caption{\label{3dspot} The color background shows the distribution of the magnetic field
on the surface of the star. The blue contours show the energy distribution in the accretion spot.
 The solid black lines show the dipole, octupole and rotational axes.}
\end{center}
\end{figure}

\begin{figure}
\begin{center}
\includegraphics[width=8.0cm]{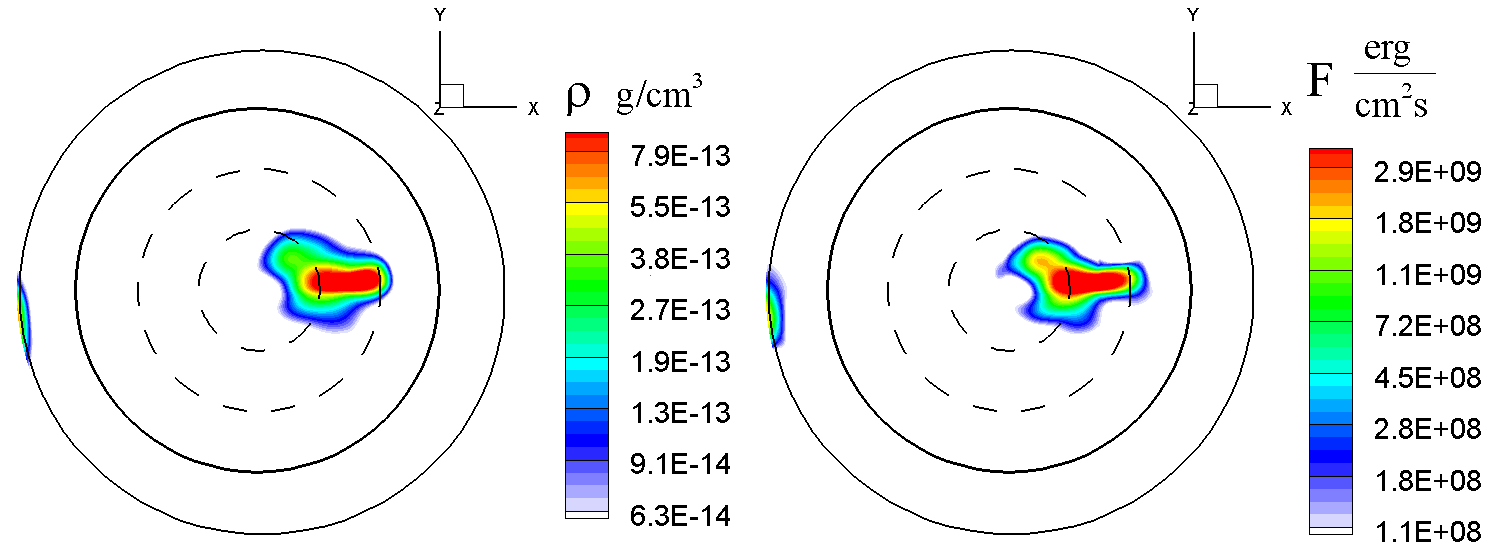}
\caption{\label{hspb} The simulated accretion spots in a
polar projection down to colatitude
$120^\circ$ from the north pole. Left panel:
density distribution; right panel: energy flux distribution.
The equator is shown as a bold line.  The dashed lines
represent the latitude $30^\circ$ and $60^\circ$ respectively.}
\end{center}
\end{figure}

D08 analyzed accretion spots from the observed brightness enhancements in chromospheric lines,
such as Ca\textsc{ii} IRT and He\textsc{i} which presumably form in (or near) the shock front close to
the stellar surface. The random flaring component,
as well as the time-variable veiling component which reflects variation of the intrinsic accretion rate were
removed from the modeling. It was also suggested that matter flows in the vicinity of the
strong magnetic field and hence Zeeman-splitting features were used for analysis.

The corresponding maps of the local surface brightness are shown in Fig. 9 of D08 for two
epochs  of observations (Feb06 and Dec06). It can be seen that in both plots there is one
bright spot located at colatitudes of $30^\circ-70^\circ$ and centered at $\theta_c\approx 45^\circ$,
while a lower-brightness area spreads up to a colatitude of $90^\circ$. In both cases the spots
are elongated in the meridional direction. The Feb06 spot also has an antipodal spot of weaker
brightness and of similar shape.  The accretion filling-factor is shown in Fig. 13 of D08.
It shows a spot located at colatitudes of $0^\circ-50^\circ$ and centered
at $\theta_c\approx 10^\circ-20^\circ$.

In our simulations the spots represent a slice of density/energy taken across the
funnel stream at the surface of the star. Hence, the spots show the distribution of these
values across the stream \citep{roma04a}. The physics of disk-magnetosphere interaction
is expected to be more complex (e.g., \citealt{kold08, cran08, cran09, bric10}). However,
this more complex physics strongly depends on the properties of funnel streams. Hence we
call these spots ``accretion spots" and show the density distribution in spots and
also the energy flux:  $F=\rho\bm{v}\cdot\hat{r}[(v^2-v_\star^2)/2+\gamma p/(\gamma-1)\rho]$
 distribution (right panel) at $t=10$, where $v_\star$ is the velocity of the star.

\fig{hsb} shows the density distribution at the surface of the star at $t=10$.
It can be seen that there are two antipodal spots  which are centered in the $\bm{\mu_1} - \bm{\Omega}$
plane slightly below the dipole magnetic pole. The spots are elongated in the meridional direction
(unlike the spots in a pure dipole configuration, see also \S 3.4).

\fig{3dspot} shows the density distribution in the spot
(contour lines)
overlaid on top of the magnetic field distribution. It can be seen that the spot is located near and below
the dipole magnetic pole and hence its position is strongly influenced by the dipole
component.  The spot is located far away from the main, high-latitude magnetic pole,
dominated by the octupolar component. We conclude that this low-latitude
position of the spot results from the fact that
the dipole field governs the matter flow, and only near the star,
the octupolar component ``comes to play" and the octupolar field
redirects the spot's position towards slightly higher latitudes, and changes the spot's shape.

\fig{hspb} shows a polar projection of the density distribution (left panel)
 and  energy flux.
It can be seen  that the accretion spot is centered near the
latitude of $40^\circ$ and spreads between latitudes of $10^\circ$
and $60^\circ$.  These spots have a close resemblance to the spots
observed by D08. In both cases, the spots are stretched in the
meridional direction and are located between latitudes of
$10^\circ$ and $60^\circ$. Analysis of our spots at different
moments of time has shown that the positions of the spots vary
only slightly with time (within 0.1 in phase). However, the spots
observed by D08 (their Fig. 9) are located at different phases.
Feb06 observations show the main spot located at a higher phase
compared with our spot, with a phase difference of 0.15-0.2.  This
is quite good agreement, considering the fact the D08 method of
the dipole moment phase reconstruction has an error in phase of
0.1-0.2 or larger.  The phase difference is higher in the Dec06
observations where the simulated and observed spots are almost in
anti-phase.

To investigate the role of the dipole and octupole phases in the
spot's position, we considered a few ``extreme" cases, where the
phase difference between the dipole and octupole moments is
$\phi=0^\circ$, $\phi=90^\circ$ and $\phi=-90^\circ$ (we have a
phase difference of $\phi=180^\circ$ in the main case). We
observed that the accretion spot is similar in shape in all these
cases, but is located at the phase corresponding to the phase of
the dipole (in the $\bm\mu_1-\bm\Omega$ plane). We conclude that
the phase of the accretion spots on the surface of the star gives
a strong indication of the dipole component's position.

\subsection{Comparison with a pure dipole model}\label{comparison}

To understand the role of the octupolar field component
 in our dipole plus octupole model of BP Tau,
we investigate a similar model but with zero octupolar field,
that is $B_{1\star}=1.2$ kG, $\Theta_1=20^\circ$ and
$B_{3\star}=0$.

\begin{figure*}
\begin{center}
\includegraphics[width=12.0cm]{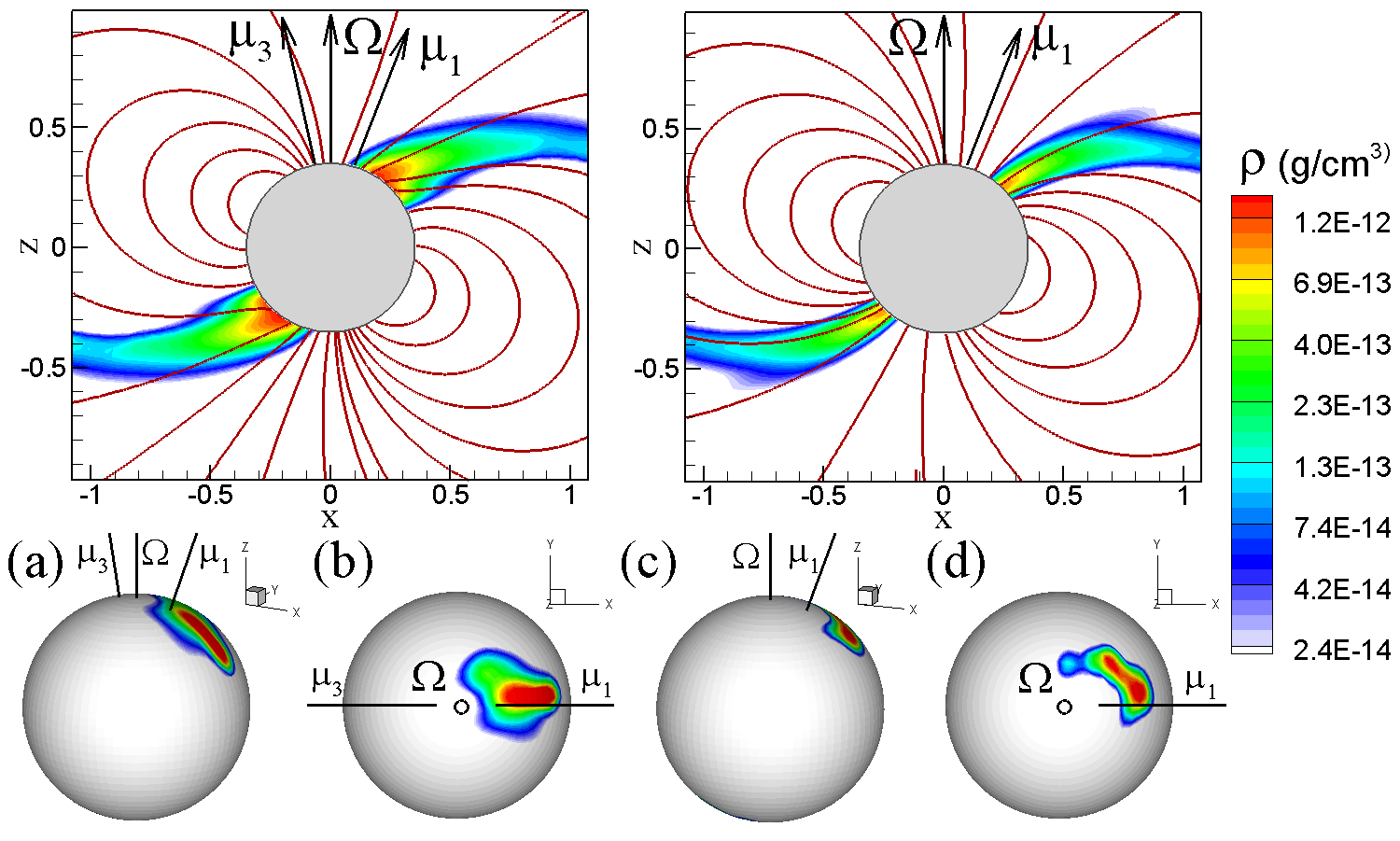}
\caption{\label{bptau_comp} Comparison of matter flow and accretion spots in the dipole plus
octuple model of BP Tau (left panels) with a pure dipole case (right panels) at $t=10$. The top panels
show the density distribution  (color background) and the sample poloidal
field lines in the $xz-$plane. The bottom panels show the density
distribution in the accretion spots as seen along the $y-$axis (panels a,c)
and along the rotational axis (panels b,d).}
\end{center}
\end{figure*}

\fig{bptau_comp} shows $xz$ slices of the accretion flow in the
dipole plus octupole model (left panel) and pure dipole model
(right panel). One can see that the flow is wider in the dipole
plus octupole model, because near the star the octupole field
influences the matter flow and redirects it in such a way that
some matter flows towards the direction of the octupolar magnetic
pole located at anti-phase with the dipole pole. The streams are
narrower in the case of the pure dipole field.

The bottom panels show that the
accretion spots in the dipole plus octupole model are meridionally
elongated, while in the dipole model they have their typical crescent shape and are elongated
in the azimuthal direction (as usually seen in the pure dipole cases, e.g. \citealt{roma03, roma04a}).

We conclude that in BP Tau, the relatively weak octupolar magnetic field which dominates only
very close to the surface of the star, strongly influences the shape and position of
the accretion spots. This influence is not as dramatic as in another T Tauri star, V2129 Oph,
which has a much weaker dipole component, and where the octupolar field
splits the funnel stream into polar and octupolar belt flows
\citep{roma10}.

\subsection{Area covered with spots}

\begin{figure}
\centering
\includegraphics[width=7.0cm]{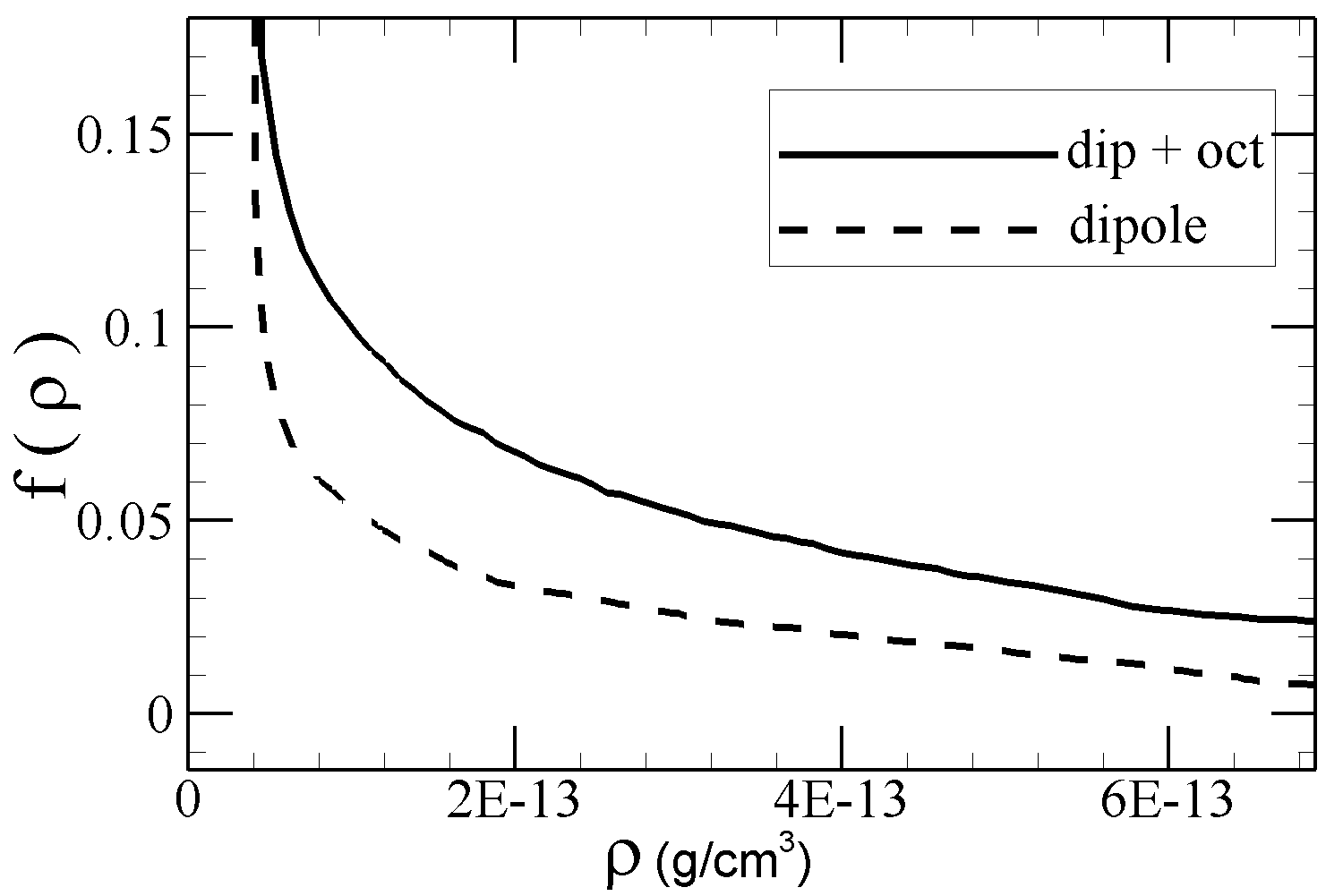}
\caption{\label{areab} Fraction of the star's surface covered by spots of density $\rho$ and higher at $t=10$.
The solid and dashed lines represent the dipole plus octupole model and dipole model respectively.}
\end{figure}

\begin{figure*}
\centering
\includegraphics[width=14.0cm]{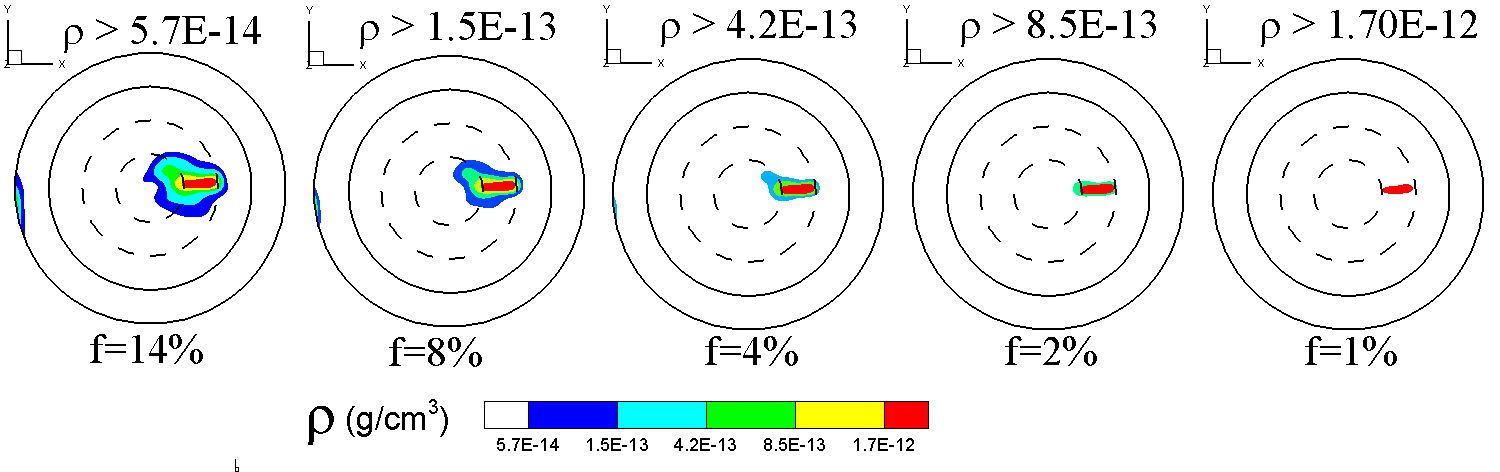}
\caption{\label{spotssize} The density distribution of spots with
a cutoff at different density levels $\rho$, and the corresponding
spot coverage  $f$ for the sum of the visible and antipodal
spots}.
\end{figure*}

D08 calculated the fraction of the star covered with visible
chromospheric spots on BP Tau and concluded that they are spread
over up to about $8\%$ of the stellar surface, and cover about
$2\%$ of the surface, assuming one-fourth of each surface pixel is
subject to accretion (see Figs. 9 \& 13 in D08). In this paper we
are interested only in the area of the spots spreading that is in
the area covered by the funnel stream, and hence we take the
larger value (8\%). Figs. 9 and 13 of D08 show that the actual
size of the spots depends on the brightness. Sometimes we take the
brightest parts of spots (4-5)\% and neglect the dimmer parts for
comparison. In observations, the spot's coverage is calculated for
one (visible)  spot versus the whole area of the star, while in
the simulations we take into account both spots, and hence the
area in simulations is twice as large compared with observations.

We observed from the simulations that the size of the accretion spot depends
on the density (or energy) cutoff. We calculate the fraction of the star covered with the spots as
 $f(\rho)=A(\rho)/4\pi R_\star^2$, where
$A(\rho)$ is the area covered with (all) spots of density $\rho$
or higher. \fig{areab} shows the distribution of $f(\rho)$ versus
$\rho$ for the dipole plus octupole model and dipole model of BP
Tau. It can be seen that for almost all chosen density levels, the
accretion spots occupy a larger surface area in the dipole plus
octupole model than in the dipole model. This result is in good
agreement with that obtained from \fig{bptau_comp}.

We choose several cutoff
densities
and show the spots in \fig{spotssize}.  One can see that in both
simulations and observations, spots cover different areas
depending on the brightness/energy flux levels ranging from $14\%$
up to $2\%$ in simulations (for two spots). In both observations
(see Fig. 9 of D08) and simulations (see our Fig.
\ref{bptau_comp}), the brightest parts of spots are located at
$30^\circ<\theta_c<60^\circ$.

It was predicted that the area covered by spots in the case of
complex magnetic fields should be smaller than in the pure dipole
case (e.g., \citealt{moha08, greg08}, see also
\citealt{calvet98}). This paper and our previous work
\citep{long08} show that the area $f$ could show a complex trend
for mixed dipole and multipole configurations, depending on how
the multipole field redirects the accretion flow.

\subsection{Accretion rate in BP Tau and in our model}

The mass accretion rate of BP Tau obtained from observations is not uniquely determined, and
different authors give different results.
 For example, \citet{gull98}
estimated the mass accretion rate to be $2.9\times10^{-8}M_\odot\mathrm{yr}^{-1}$  using the
luminosity of $L\simeq (GM_\star\dot{M}/{R_\star})(1-R_\star/R_{in})$,
 where $R_{in}$
is the inner radius of the disk.  Other estimates of the mass accretion rate include
$1.6\times10^{-7}M_\odot\mathrm{yr}^{-1}$ \citep{valen04},
$9\times10^{-10}M_\odot\mathrm{yr}^{-1}$ \citep{schm05}. \citet{calvet04} compared
the mass accretion rates of intermediate-mass,  $(1.5-4)M_\odot$, and
low-mass, $(0.1-1.0)M_\odot$, T Tauri stars and  concluded that the average mass
accretion rate for intermediate-mass T Tauri stars is about $3.0\times10^{-8}M_\odot\mathrm{yr}^{-1}$,
while for low-mass CTTSs (like BP Tau) it is about
5 times smaller, that is, $6.0\times10^{-9}M_\odot\mathrm{yr}^{-1}$.

We obtained from simulations the dimensionless accretion rate of $\widetilde{\dot M}\approx 0.13$.
Using the reference value for $\dot M_0$ from Tab. \ref{tab:refval},
 $\dot M_0=1.1\times10^{-8}M_\odot\mathrm{yr}^{-1}$
(calculated for $\widetilde\mu_1=3$), we obtain the dimensional
accretion rate: $\dot M=\dot M_0 \widetilde{\dot M}\approx
1.4\times 10^{-9}M_\odot\mathrm{yr}^{-1}$. This value  is higher
than the accretion rate suggested by \citet{schm05} but lower than
the other estimates. Given that BP Tau has the low mass of $\sim
0.7M_\odot$ (D08) (or $0.49M_\odot$ \citealt{gull98}), the mass
accretion rate from our simulation models may still be in the
reasonable range according to the \citet{calvet04} analysis and
discussion.

Here, we should note that the ``main model" considered above is one of the suggested models, where the disk
is truncated at large distances. Below, we discuss additional models, where the
disk is truncated at smaller distances and the accretion rate is higher.

\subsection{Modeling of BP Tau at higher accretion rates}

We performed two additional simulation runs at higher accretion rates (at smaller values of
parameter $\widetilde\mu_1$) keeping the ratio between the dipole and octupole moments fixed,
  $\widetilde\mu_3/\widetilde\mu_1=0.08$.
The parameters of these new models are the following:
$\widetilde\mu_1=2, \widetilde\mu_3=0.16$ and
$\widetilde\mu_1=1, \widetilde\mu_3=0.08$.

\begin{figure}
\centering
\includegraphics[width=8.5cm]{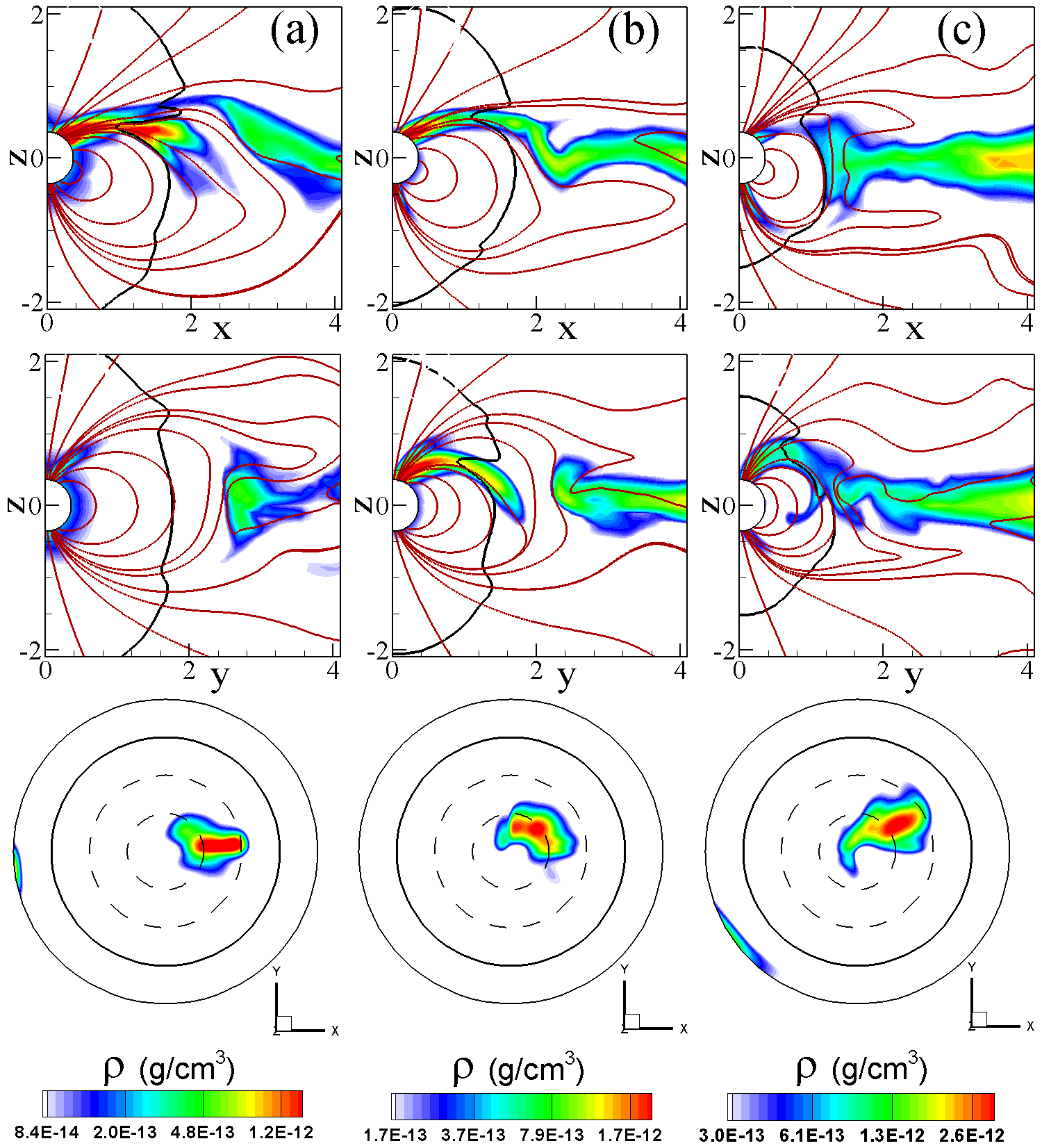}
\caption{\label{compareb} Simulations at different accretion rates
at $t=8$. (a) main case, $\widetilde\mu_1=3,
\widetilde\mu_3=0.24$; (b) $\widetilde\mu_1=2,
\widetilde\mu_3=0.16$; (c), $\widetilde\mu_1=1,
\widetilde\mu_3=0.08$. The top panels show the density
distribution (color background), the poloidal field lines (red
lines) and the $\beta=1$
 line (gray) in the $xz-$plane. The bottom panels show the energy flux
 distribution in the accretion spots
 in a polar projection.}
\end{figure}

\fig{compareb} (middle and right panels) shows the matter flow and
accretion spots for these two cases, while the left panel shows
results for the main model ($\widetilde\mu_1=3,
\widetilde\mu_3=0.24$) for comparison. One can see that in the new
models, the disk is truncated at smaller distances from the star:
$r_t\approx(5-6) R_\star$ (for $\widetilde\mu_1=2$) and
$r_t\approx3.6 R_\star$ (for $\widetilde\mu_1=1$). The bottom
panels of \fig{compareb} show that when the disk comes closer to
the star, the spots are still at high latitudes, though they
become longer in the meridional direction. This is probably
because the octupolar component has a stronger influence on the
spots: the spots move closer to the octupolar magnetic pole. Note
that no octupolar ring spots appear (like in V2129 Oph, see R10).
This is probably because the disk truncation radius is still far
away from the area where octupole has a strong influence on the
flow, which is closer to the star.  It is interesting to note that
in the pure dipole case, the spots  move towards lower latitudes
when the disk comes closer to the star. Here, we see the opposite:
parts of the spots move to higher latitudes, because the octupolar
component becomes more significant at smaller truncation radii.

The dimensionless mass accretion rate which we obtain from our
simulations is $\widetilde{\dot M}\approx 0.075$ (for
$\widetilde\mu_1=2$) and $\widetilde{\dot M}\approx 0.085$ (for
$\widetilde\mu_1=1$). We take reference values $\dot M_0$ for
different $\widetilde\mu_1$ from Tab. \ref{tab:refval} and obtain
corresponding dimensional accretion rates $\dot M$ (see Tab.
\ref{tab:mdot-angmom}) which are  $2.5\e{-9}M_\odot$yr$^{-1}$ and
$8.5\e{-9}M_\odot$yr$^{-1}$ respectively. We see that the
accretion rate for $\widetilde\mu_1=1$ is high and close to many
of the  observed values. However, the disk comes too close to the
star and the truncation radius is much smaller than the corotation
radius ($R_{cor}\approx 7.5 R_\star$), which would mean that the
star is not in the rotational equilibrium state, which is
unlikely. The model with $\widetilde\mu_1=2$ is similar to the
main model: the disk is truncated far away and the accretion rate
is twice as high, but still it is lower than many values given by
observations. This model is slightly better than the ``main"
model.

In the case of a pure dipole, we obtain from simulations
$\widetilde{\dot M}\approx 0.12$ and $\dot M\approx
1.3\e{-9}M_\odot$yr$^{-1}$. \ref{tab:mdot-angmom} summarizes the
results obtained in our simulation models and some observational
properties of BP Tau for comparison.

\begin{table*}
\caption{Comparison of observational and modeled results shown in
D08 and results from our simulations.} \centering
\begin{tabular}{lllllll}
\hline
Models                            &  $r_t$    & $\widetilde{\dot{M}}$ & $\dot{M}$($M_\odot$yr$^{-1}$) & $\widetilde{N_f}$ & $N_f$ (g cm$^2$s$^{-2}$) & $\tau$ (yr)      \\
\hline
D08 \& other measurements                                 &  $>4R_\star$          & --        & $10^{-9} - 10^{-7}$ & --& \\
Model  ($\widetilde\mu_1=3, \widetilde\mu_3=0.24$)        &  $(6-7)R_\star$       & $0.13$     & $1.5\e{-9}$ & $-0.03\pm0.014$ & $-1.9\e{35} $ & $1.6\e7$ (spin-down)  \\
Model ($\widetilde\mu_1=2, \widetilde\mu_3=0.16$)         &  $(5.4-6.4)R_\star$   & $0.075$   & $1.9\e{-9}$ & $-0.005\pm0.01$ & $-1.5\e{35}$ & $2.1\e7$ (spin-down) \\
Model ($\widetilde\mu_1=1, \widetilde\mu_3=0.08$)         &  $3.6 R_\star$     & $0.075$   & $7.7\e{-9}$ & $+0.002\pm0.004$ & +$2.3\e{35}$ & $1.1\e7$ (spin-up) \\
Model ($\widetilde\mu_1=3, \widetilde\mu_3=0$)            &  $(6-7)R_\star$       & $0.12$    & $1.4\e{-9}$ & $-0.045\pm0.015$ & $-2.4\e{35}$ &  $1.3\e7$ (spin-down)  \\
\hline
\end{tabular}
\label{tab:mdot-angmom}
\end{table*}

\subsection{Angular momentum transport}

We calculated torque on the surface of the star for the above
three models. The torque associated with the magnetic field
$\tilde{N_f}$ is about 10-30 times larger than the torque
associated with matter and thus dominates (see also
\citealt{roma02,roma04a,long07,long08}). The dimensionless torque
obtained from simulations, $\widetilde N_0$, varies in time around
some average value. Tab. \ref{tab:mdot-angmom} shows these values
and values corresponding to these variations. We observed in
simulations that the torque is negative in most of the models
because the disk is truncated at the distances comparable with the
corotation radius and therefore the star spins-down. In a model
with $\widetilde\mu_1=1$, the disk is truncated at the distance of
$\sim 0.5 R_{cor}$ and it spins the star up. To estimate the
dimensional torque, $N_f$, we take the largest absolute value of
$\widetilde N_f$ from Tab. \ref{tab:mdot-angmom} and take into
account the reference values of $N_0$ from Tab. \ref{tab:refval}.
Tab. \ref{tab:mdot-angmom} shows the dimensional torque.

We estimate the time-scale of spinning-down of BP Tau. The star's
angular velocity is (at period $P=7.6$ days) $\Omega=2\pi/P
\approx 9.6\times 10^{-6} {\rm s}^{-1}$, its angular momentum is
$J= k M_\star R_\star^2 \Omega = 2.5\times  10^{50} k ~{\rm g
cm}^2/{\rm s}$, where $k<1$ (we take $k=0.4$ for estimations). The
spin-down time-scale, $\tau = J/N_f$, was estimated for different
models and is shown in Tab. \ref{tab:mdot-angmom}. One can see
that the spin-up/down time-scale is about an order magnitude
larger than the age of BP Tau  ($ 1.5\times 10^6$ years, see
references in D08). Hence, the torque obtained in simulations is
not sufficient to regulate the spin of the star at the considered
epoch. It is possible that a star lost most of its angular
momentum at the earlier stages of its evolution. Note that similar
low-torque situation has been observed in modeling of another
CTTS, V2129 Oph.


\section{Conclusions and Discussions}

We performed global 3D MHD simulations of accretion onto a model star with parameters
close to those of the classical T Tauri star BP Tau, and with the magnetic field
approximated with a superposition of slightly tilted dipole and octupole moments with polar magnetic fields of
1.2kG and 1.6kG which are in anti-phase
(D08). In this star
the dipole field dominates almost up to the surface of the star and determines the majority
of the observational properties.

We performed a number of simulation runs for different truncation
radii of the disk, and chose one of them where the disk is
truncated at $r\approx (6-7) R_\star$ which is sufficiently close
to the corotation radius $R_{cor}\approx7.5 R_\star$ and
investigated this case in greater detail. We observed that the
dipole component of the field truncates the disk, and matter flows
in two funnel streams towards the dipolar magnetic poles. However,
near the star the flow is slightly redirected  by the octupolar
component towards higher latitudes, and this affects  the shape
and position of the accretion spots: the spots are stretched in
the meridional direction and are centered at higher latitudes
compared with spots in the pure dipole case, which are
latitudinally-elongated (crescent-shaped) and are centered at
lower latitudes.

The spots are located near the $\bm{\mu_1}-\bm{\Omega}$ plane, where both  the
dipole and octupole moments are
situated, and are in anti-phase. Experiments with different relative phases between the dipole and octupole moments
have shown that the spots are always located in the meridional plane of the dipole moment.

The spot's position slightly varies in phase with the accretion rate, but
the variation is small, about $8^\circ$  at the most.
Note that in the pure dipole case, the spot may rotate about the magnetic pole
(and hence may strongly change the phase) if the dipole is only slightly tilted about the
rotational axis, $\Theta_1\lesssim 5^\circ$ (\citealt{roma03, roma04a, bace10}).
However, in BP Tau, the tilt is larger, $\Theta_1\approx 20^\circ$ and an octupolar component
is present and is significant near the star. Both factors lead to the restriction of such a motion.

The meridional position and shape of spots observed in simulations is similar to those observed
by  \citet{dona08} (see their Fig. 9 for brightness distribution in spots
reconstructed for the Dec06 or Feb06 epochs). However, they are at different phases compared with
our spots.
This may be connected with the relatively low accuracy of the phase
reconstruction of the dipole component from the surface magnetic field, or
some other reason.

The accretion rates obtained in different models are in the range
of   $\dot M\approx (1.5-7.7)\times 10^{-9}M_\odot$yr$^{-1}$.  It
is lower than most of the $\dot M$ values derived from
observations which vary in the range of $9\e{-10}$ and $1.6\e{-7}$
solar mass per year and depend on the approaches used for the
derivation. The smallest accretion rate obtained in simulations
corresponds to the case where the disk is truncated at $r_t\sim
R_{cor}$, and a star slightly spins down, while the largest to the
case where $r_t\sim 0.5 R_{cor}$ and a star spins up. We should
note that at larger accretion rates, say at   $1.6\e{-7}$ solar
mass per year, the disk will come very close to the star, and the
torque would be much stronger. Such a state seems unlikely. Hence,
if the accretion rate is very high, then we should suggest that
the dipole component should be larger than that derived by D08.



The torque obtained in simulations is  small and the time-scale of
the spinning-up/down is an order of magnitude smaller compared
with the age of BP Tau. This torque is not sufficient to support a
star in the rotational equilibrium state, where a star spins up or
spins down depending on the accretion rate but has a zero torque
on average (e.g., \citealt{ghosh79a}, \citealt{konigl91},
\citealt{camer93}, \citealt{long05}).
 We assume that this small
torque matches a small accretion rate obtained in simulations
because  the torque generally correlates with the accretion rate
(e.g., \citealt{roma02}).

We suggest that a star may lose the majority of its angular
momentum at earlier stages of its evolution due to the
``propeller" effect (e.g., Romanova et al. 2005; Ustyugova et al.
2006), stellar winds (\citealt{matt05}, 2008), or by some other
mechanism.

Earlier, we  performed global 3D MHD simulations of accretion onto V2129 Oph \citep{roma10}
which have shown that in the case of a strong octupolar component, parts of the octupolar
belt spots can be visible and can dominate at sufficiently high accretion rates.
This is not the case in BP Tau,
where the octupolar component dominates only in the close vicinity of the star.

Disk-magnetosphere interaction leads to inflation of the external field lines
and  formation of a magnetic tower.
Simulations show that the potential approximation used in extrapolation of the magnetic
field from the surface of the star to larger distances (e.g. \citealt{jard06}, D08, \citealt{greg10}) is
valid only inside the magnetospheric (Alfv\'en) surface,  where the magnetic stress
dominates. At larger distances, the magnetic field distribution
strongly departs from potential.




\section*{Acknowledgments}

Resources supporting this work were provided by the NASA High-End Computing (HEC)
Program through the NASA Advanced Supercomputing (NAS) Division at Ames Research Center and
the NASA Center for Computational Sciences (NCCS) at Goddard Space Flight Center.
 The authors thank A.V. Koldoba and G.V. Ustyugova for the earlier development of
the codes.  The research was supported by NSF grant AST0709015. The research of MMR was supported by NASA grant
NNX08AH25G and NSF grant AST-0807129.


\end{document}